\documentclass[aps,prl,twocolumn,superscriptaddress]{revtex4-1}
\usepackage{bm}
\usepackage{graphicx}
\usepackage{color}
\usepackage{hyperref}

\makeatletter

\newcommand{\Rmnum}[1]{\expandafter\@slowromancap\romannumeral #1@}
\usepackage{amsmath}

\begin{document}

\preprint{APS/123-QED}

\title{Surface band-selective moiré effect induces flat band in mixed-dimensional heterostructures}% Force line breaks with \\

\author{Shuming Yu}
 \affiliation{Institute for Advanced Studies, Wuhan University, Wuhan 430072, China}
\author{Zhentao Fu}
 \affiliation{Institute for Quantum Science and Technology, Shanghai University, Shanghai 200444, China}
\author{Dingkun Qin}
 \affiliation{Institute for Advanced Studies, Wuhan University, Wuhan 430072, China}
\author{Enting Li}
 \affiliation{Institute for Advanced Studies, Wuhan University, Wuhan 430072, China}
\author{Hao Zhong}
 \affiliation{Institute for Advanced Studies, Wuhan University, Wuhan 430072, China}
\author{Xingzhe Wang}
 \affiliation{Institute for Advanced Studies, Wuhan University, Wuhan 430072, China}
\author{Keming Zhao}
 \affiliation{Institute for Advanced Studies, Wuhan University, Wuhan 430072, China}
\author{Shangkun Mo}
 \affiliation{Institute for Advanced Studies, Wuhan University, Wuhan 430072, China}
\author{Qiang Wan}
 \affiliation{Institute for Advanced Studies, Wuhan University, Wuhan 430072, China}
\author{Yiwei Li}
 \affiliation{Institute for Advanced Studies, Wuhan University, Wuhan 430072, China}
 \author{Jie Li}
 \affiliation{Institute for Quantum Science and Technology, Shanghai University, Shanghai 200444, China}
\author{Jianxin Zhong}
 \affiliation{Institute for Quantum Science and Technology, Shanghai University, Shanghai 200444, China}
\author{Hong Ding}
 \affiliation{Tsung-Dao Lee Institute and School of Physics and Astronomy, Shanghai Jiao Tong University, Shanghai 200240, China}
\author{Nan Xu}
 \email{nxu@whu.edu.cn}
 \affiliation{Institute for Advanced Studies, Wuhan University, Wuhan 430072, China}%

\date{\today}% It is always \today, today,

\begin{abstract}
In this work, we reveal a curious type of moiré effect that selectively modifies the surface states of bulk crystal. We synthesize mixed-dimensional heterostructures consisting of a noble gas monolayer grow on the surface of bulk Bi(111), and determined the electronic structure of the heterostructures using angle-resolved photoemission spectroscopy . We directly observe moiré replicas of the Bi(111) surface states, while the bulk states remaine barely changed. Meanwhile, we achieve control over the moiré period in the range of 25 Å to 80 Å by selecting monolayers of different noble gases and adjusting the annealing temperature. At large moiré periods, we observe hybridization between the surface band replicas, which leads to the formation of a correlated flat band. Our results serve as a bridge for understanding the moiré modulation effect from 2D to 3D systems, and provide a feasible approach for the realization of correlated phenomena through the engineering of surface states via moiré effects.

%\begin{description}
%\end{description}
\end{abstract}

\maketitle

%\tableofcontents

%\section{Introduction}

The peculiar properties of crystals are dependent on their distinct periodic structures. Notably, the modulation of the periodic structure in two-dimensional (2D) systems using moiré superlattices has garnered significant interest\cite{doi:10.1073/pnas.1108174108,cao2018unconventional,cao2018correlated,Zhang:2020aa,chen2020tunable}. By adjusting the twist angle between two-dimensional (2D) layers, moiré band hybridizations induce flat bands with narrow bandwidth, which further gives rise to novel correlated states, including the correlated insulator\cite{cao2018correlated,Cao:2020aa}, unconventional superconductivity\cite{cao2018unconventional,doi:10.1126/science.aav1910,Lu:2019aa}, general Wigner crystal\cite{Regan:2020aa,Padhi:2018aa}, Rydberg moiré excitons\cite{doi:10.1126/science.adh1506} and fraction quantum anomalous Hall effect\cite{doi:10.1126/science.aay5533}. Moreover, high-order moiré effect has been realized in systems with large lattice constant mismatch\cite{PhysRevLett.129.176402, WOS:001283292600008, https://doi.org/10.1002/adma.202305175, PhysRevB.104.235130}.The moiré period can be in situ tuned in an alternative way, by controlling lattice constant of noble gas monolayer\cite{PhysRevLett.129.176402, PhysRevB.109.L161102, PhysRevB.110.085148}. 

\begin{figure}[htbp]
	\centering
	\includegraphics[width=1\columnwidth]{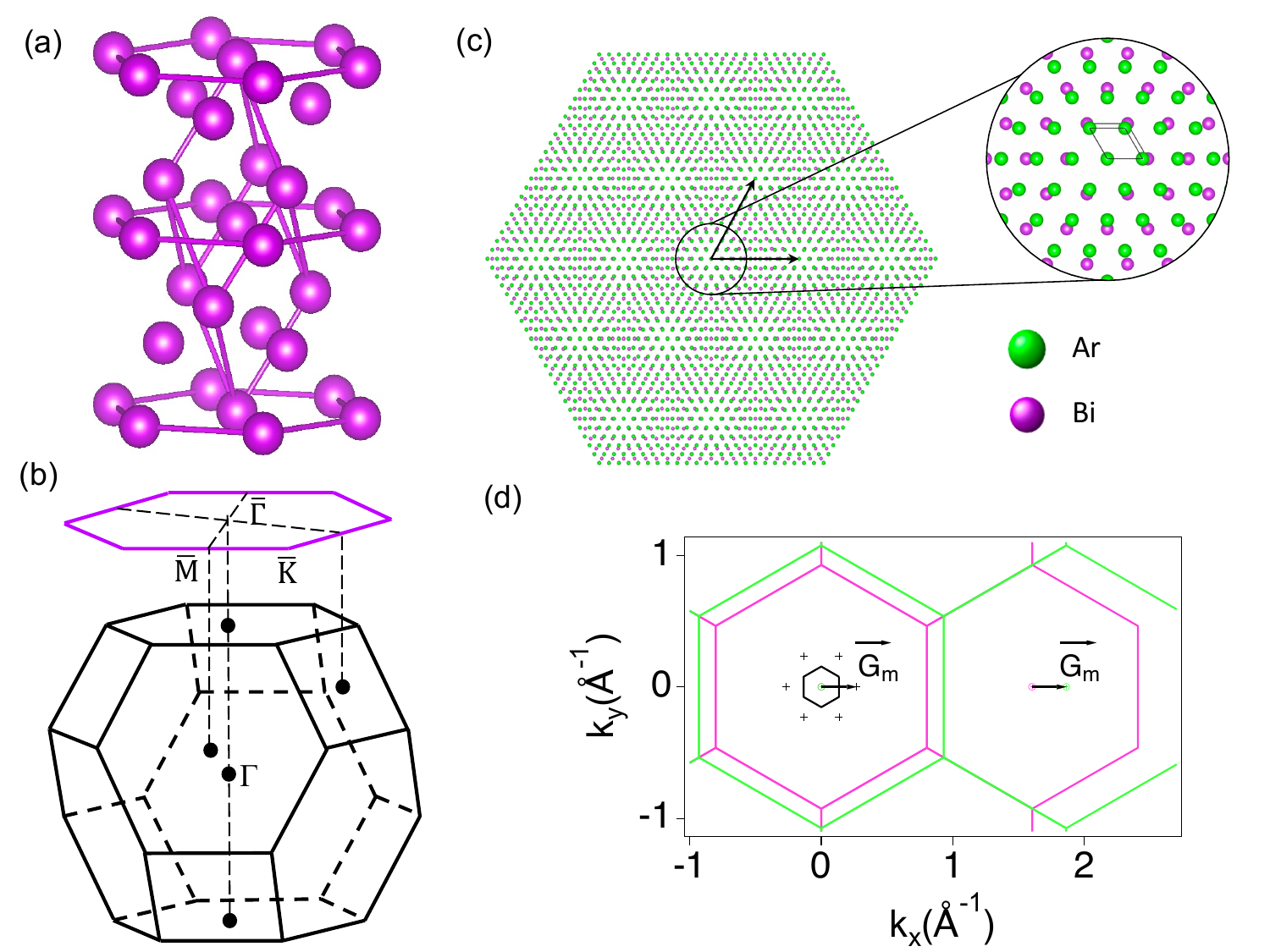}
	\caption{(a) Crystal structure of bismuth, the inside rhombohedron is the primitive unit cell and the outside hexagonal is the conventional cell. (b)  Bulk Brillouin zone(black) and surface Brillouin zone(pink) of bismuth.(c) Long-range moiré pattern of mAr/Bi in real space, right picture is the center zoomed-in area. (d) Moiré effect of Bi/mAr, the pink/green/black line is Bi/Ar/moiré Brillouin zone }
\end{figure}

Interestingly, mixed-dimensional moiré effect has been discovered in graphite-based heterostructure\cite{PhysRevLett.129.176402,PhysRevB.109.L161102} and $Cu_{x}TiSe_{2}$/noble gas monolayer heterostructure\cite{PhysRevB.110.085148}. In mixed-dimensional systems, the interfacial moiré effect exerts a global modification on the bulk electronic structure of semimetals, thereby expanding the scope of moiré-modulated objects from two-dimensional to three-dimensional (3D) systems. The achievement of the transitional state from the 2D moiré effect to the 3D mixed-dimensional moiré effect is not only crucial for deepening the understanding of the moiré effect from a dimensional perspective, but also conducive to expanding the applications of the moiré effect through the engineering of electronic states.

\begin{figure*}[htbp]
	\centering
	\includegraphics[width=1.5\columnwidth]{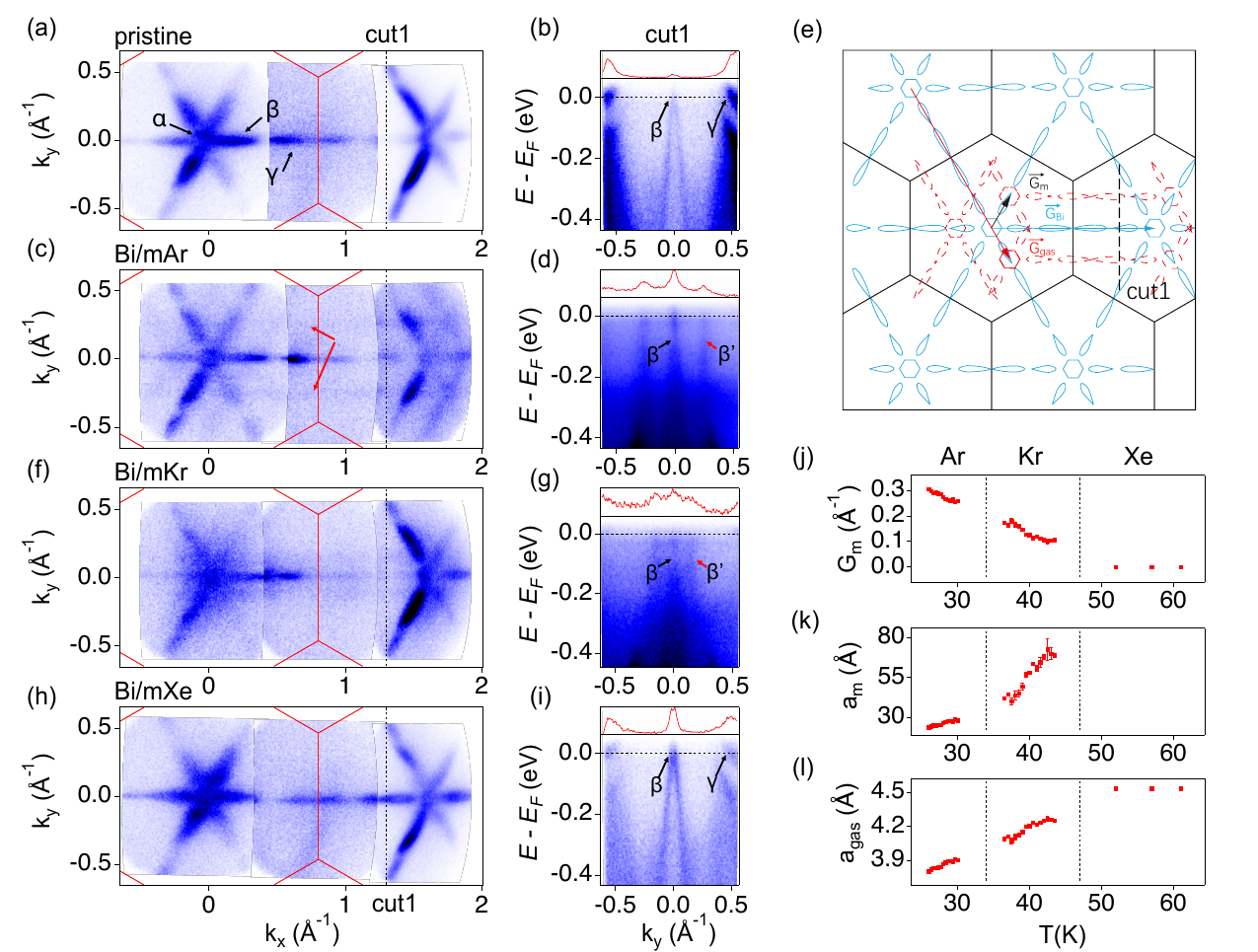}
	\caption{ (a) Fermi surface mapping of Bi(111) along $\bar{\Gamma}$ - $\bar{M}$ - $\bar{\Gamma}$ direction. (b) Band structure measured along $k_{x}=1.3Å^{-1}$ in (a), red curve is MDC at fermi level.  (c, d) Same as (a, b) but measure from monolayer Ar and Bi(111) heterostructure, the red arrows point to the moiré replicas. (e) Schematic moiré effect in noble gas monolayer on Bi(111) surface. (f-i) Same as (a, b) but measured from monolayer Kr/Xe and Bi(111) heterostructure. (j, k, l) Temperature dependence of $\Delta\Gamma$, $a_{m}$, $a_{gas}$. }
\end{figure*}

In this letter, we report a surface band-selective moiré effect in a mixed-dimensional heterostructure, which is achieved via the in situ deposition of a noble gas monolayer on the (111) surface of a bismuth (Bi) single crystal. By employing angle-resolved photoemission spectroscopy (ARPES), we observe that the interfacial moiré effect selectively modulates the surface states of Bi(111) and induces the formation of moiré replicas, while leaving the bulk states unaltered. The moiré period can be tuned over a wide range by depositing different noble gas monolayers and performing annealing treatments at various temperatures. Specifically, under the condition of a large moiré period, we observe that the surface states and their replicas exhibit mutual hybridization and further form flat bands. Our results not only provide the transitional state of moiré modulation effect from 2D to 3D, but also uncover a novel method to realize the correlated phenomena by moiré engineering surface state.

\begin{figure}[htbp]
	\centering
	\includegraphics[width=1\columnwidth]{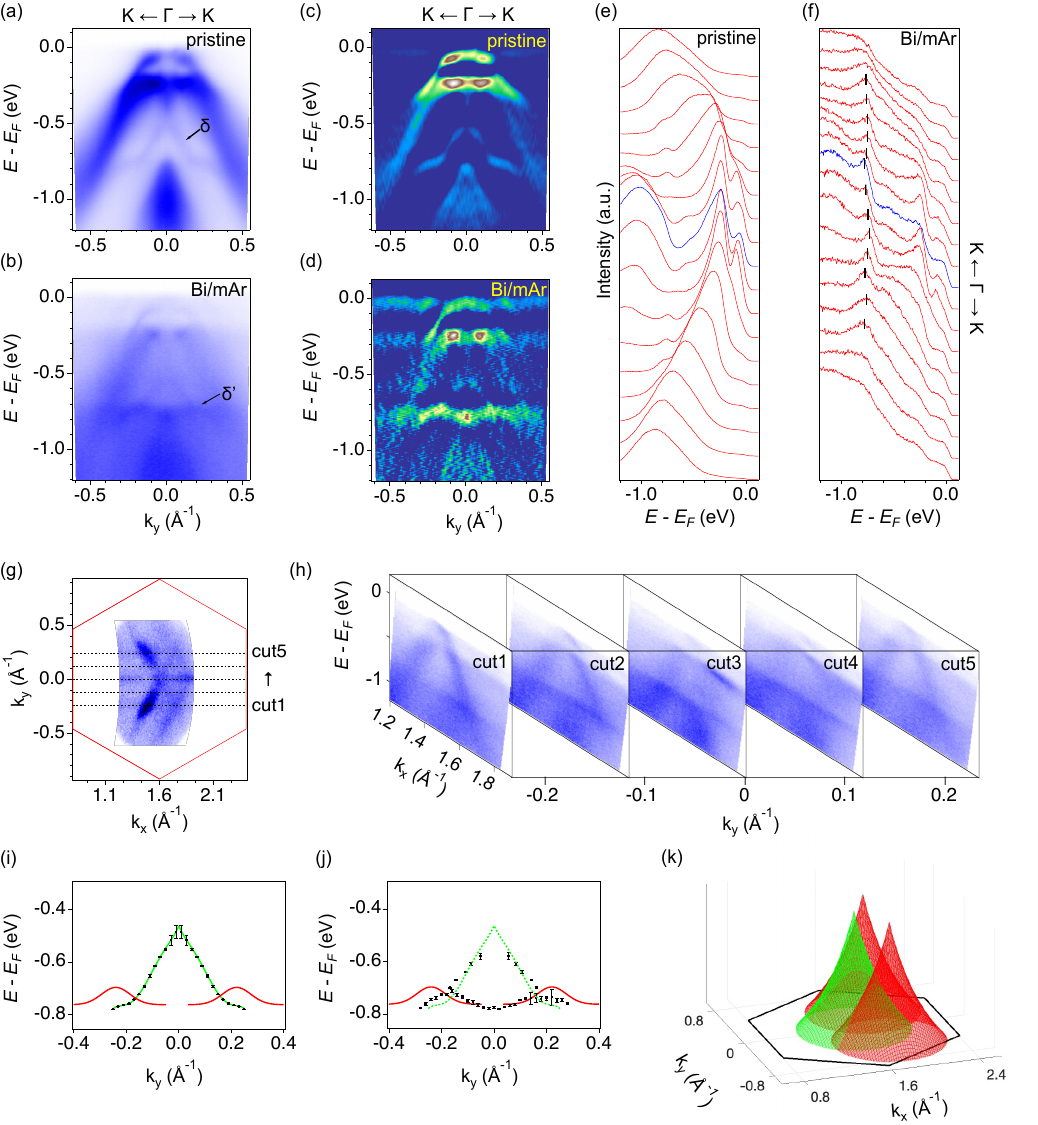}
	\caption{ (a) Pristine Bi band structure measured along $\Gamma$-K direction in second Brillouin zone.  (b) Bi(111)/mAr band structure measured along $\bar{\Gamma}-\bar{K}$ direction in second Brillouin zone, the moiré effect flat band $\delta’$ is point by black arrow.  (c, d) Curvature of (a, b). (e, f) EDC from band structure in $\Gamma$-K direction with and without monolayer Ar, the black dash line indicates flat band $\delta’$. (g) Fermi surface mapping of Bi(111)/mAr in second Brillouin zone. (h) Typical band structure in (g) at $k_{y}$=-0.24$Å^{-1}$ (cut1), -0.12$Å^{-1}$ (cut2), 0$Å^{-1}$ (cut3), 0.12$Å^{-1}$ (cut4), 0.24$Å^{-1}$ (cut5). (i, j) Evolution of band $\delta$, black dots are EDC peak fitting,  green line is dots fitting of band $\delta$ and $\delta’$, red line is theoretical moiré band $\delta_{m}$ curve at $\Gamma - K$ direction. (k)Schematic of pristine and moiré replicas of band $\delta$, green part is pristine band $\delta$ and red part is theoretical moiré band  $\delta_{m}$.}
\end{figure}

ARPES experiments are carried out at the lab-based ARPES system using helium lamp ($\hbar\nu$ = 21.2 eV). Bi single crystals are cleaved in situ in a base vacuum better than $1 \times10^{-10}$ mbar, to obtain a clean and flat Bi(111) surface. Noble gas monolayers are deposited on fresh surfaces with gas partial pressure below $2 \dim 10^{-9}$ mbar. The number of noble gas layers is confirmed by ARPES results on valence band structures of noble gas layers at high binding energy ($E_{B}$). 

Bi crystallizes in a rhombohedral structure, where the Bi(111) surface serves as its natural cleavage plane (Fig. 1a). Each Bi(111) bilayer consists of a dislocated stacking arrangement with a triangular structure, and the corresponding surface Brillouin zone (BZ) is presented in Fig. 1b.The bulk Bi crystal exhibits a semimetallic behavior and possesses strong spin-orbit coupling (SOC) \cite{PhysRevLett.93.046403,HOFMANN2006191}. The Bi(111) surface hosts characteristic surface states with Rashba splitting, which are well separated from the projection of bulk bands \cite{HIRAHARA201598}. Through isovalent antimony (Sb) substitution, a 3D topological insulator phase is achieved in the $Bi_{1-x}Sb_{x}$ alloy system \cite{Hsieh:2008aa,PhysRevB.78.045426}, which is accompanied by topological surface Dirac states.

In Fig. 1c, we present the simulation results of the large-scale atomic arrangement of a noble gas monolayer adsorbed on the Bi(111) surface, where the orientation of the monolayer is verified by the ARPES results of valence bands. The moiré pattern is readily identifiable, which arises from the small lattice mismatch between the Bi(111) surface and the noble gas monolayer. The primitive vectors of the moiré reciprocal lattice and the moiré BZ are determined by the difference between the reciprocal primitive vectors of the Bi(111) surface and those of the noble gas monolayer, following the relation  $\vec{G}_{m} = \vec{G}_{Bi(111)}-\vec{G}_{gas}$ (Fig. 1d).

The band structures of the Bi(111)/noble gas monolayer heterostructures are presented in Fig. 2. The Fermi surface (FS) of pristine Bi(111) consists of three types of pockets, which are formed by a pair of surface states with Rashba splitting (Fig. 2a). Specifically, these pockets include: a small electron pocket $\alpha$ centered at the $\Gamma$ point; petal-shaped hole pockets $\beta$ along the $\Gamma$-M high-symmetry lines (Fig. 2b); and another set of petal-shaped electron pockets $\gamma$ near the M point. For the Bi(111)/argon monolayer (Bi/mAr) heterostructure, FS clearly exhibits replicas of surface states induced by the moiré pattern (Fig. 2c). These surface state replicas can be better resolved in the band structure plot along Cut 1, as shown in Fig. 2d. The mechanism underlying the formation of these replicas is illustrated in Fig. 2e: the moiré potential causes a shift in the original surface states, with the shift magnitude determined by the reciprocal primitive vector of the moiré lattice $\vec{G}_{m}$. 

Figs. 2f–g present the ARPES results of the FS and band structure along cut 1 for the Bi(111)/krypton monolayer (Bi/mKr) heterostructure. In Bi(111)/mKr, the replicas of surface states are closer to each other, which indicates a larger moiré period $a_{m}$; this observation is consistent with the smaller lattice mismatch between Bi(111) and Kr. Specifically, the smaller lattice mismatch in Bi/mKr gives rise to a larger $a_{m}$, which in turn corresponds to a smaller moiré reciprocal lattice vector $\vec{G}_{m}$.

In the Bi(111)/xenon monolayer (Bi/mXe) heterostructure, the lattice constant of the Xe monolayer is pinned by the Bi(111) surface, resulting in $a_{Xe} = a_{Bi}$. Consequently, no moiré pattern is observed in Bi/mXe, and there are no replica bands present (Figs. 2h–i). On the other hand, the Xe monolayer still exerts a notable effect on Bi(111), specifically manifesting as a spectral weight redistribution among the surface states in Bi/mXe (Figs. 2h–i) when compared to that in pristine Bi(111) (Figs. 2a–b). A clear comparison of the Fermi surface momentum distribution curves (MDC) in Figs. 2b and 2i reveals the following: in pristine Bi(111), the intensity of the $\delta$ band is significantly stronger than that of the $\beta$ band; however, in Bi/mXe, the $\delta$ band becomes weaker than the $\beta$ band.

Building on our previous studies\cite{PhysRevLett.129.176402,PhysRevB.110.085148}, we tune the lattice constants of the Ar and Kr monolayers by annealing at different temperatures $T_{a}$, with the aim of varying the moiré period $a_{m}$. By measuring the $T_{a}$-dependent separation along the $k_{y}$ direction between replica bands, we can determine the $T_{a}$-dependent $\vec{G}_{m}$ using the geometry configuration illustrated in Fig. 2e. Subsequently, the $T_{a}$-dependent $a_{m}$ and $a_{gas}$ can be derived from this result. The annealing temperature $T_{a}$-dependent results are summarized in Figs. 2j-l. In Bi/mAr, the moiré period can be tuned from 23.6 Å to 28 Å; in Bi/mKr, this range extends from 41.8 Å to 72.8 Å. Additionally, the $T_{a}$-dependent $a_{gas}$ values for Bi(111)/mAr and Bi(111)/mKr are in good agreement with our previous findings\cite{PhysRevLett.129.176402,PhysRevB.110.085148}.

The results presented in Fig. 2 clearly illustrate the moiré modulation of Fermi pockets in the surface states of Bi(111)/noble gas monolayer heterostructures. To further investigate the moiré effect on bulk states and the electron interactions between pristine bands and moiré replicas, we conducte a comprehensive study of the high-resolution band structure along high-symmetry directions. In Figs. 3a and 3b, we directly compare the band structures of pristine Bi(111) and Bi/mAr (with a moiré period $a_{m}$ = 25 Å) along the $\Gamma$-K direction. The corresponding second derivative plots and energy density curve (EDC) plots are presented in Figs. 3c–d and 3e–f, respectively. Consistent with the FS results in Figs. 2c and 3g, additional surface state replicas (corresponding to the $\beta$’ -band) are observed near the Fermi level ($E_{F}$). Furthermore, a flat band emerges at the binding energy $E_{B}$ =0.75eV in Bi/mAr (Fig. 3b), which is more clearly visible in the second derivative (Fig. 3d) and EDC (Fig. 3f) plots. The bandwidth of this flat band is estimated to be within 20 meV, determined by tracing EDC peaks. This band exhibits flatness over a large momentum range, as evident from the series of band structure plots in Fig. 3h, where the momentum paths along cut 1 to cut 5 are illustrated.

The flat band observed herein is a direct consequence of moiré modulation on surface states. Consistent with previous theoretical and experimental studies\cite{PhysRevB.99.195419,PhysRevResearch.2.043127,Lisi:2021aa,Li:2024aa}, we detect additional surface states at high $E_{B}$ – labeled as $\delta$ – on the pristine Bi(111) surface, in addition to the surface states near the Fermi level $E_{F}$ (Fig. 3a). In Bi/mAr, moiré potential induces replicas bands $\delta_m$, as illustrated in Fig. 3i; this follows the same mechanism that generates the moiré replicas near the $E_{F}$ (Fig. 2e). In Fig. 3i, we trace the original $\delta$ band and simulate the replica band $\delta_{m}$ according to the momentum locations. Fig. 3j displays the experimentally determined band structure of Bi(111)/mAr, with the simulation appended. The moiré bands in Bi(111)/mAr show good agreement with the simulation in Fig. 3i, alongside a moderate hybridization effect at the band crossings. In contrast to surface states, which are strongly modulated by the moiré effect, bulk states exhibit regular behavior upon deposition of an Ar monolayer on the Bi(111) surface, characterized by diminished spectral weight and an absence of replicas.

\begin{figure}[htbp]
	\centering
	\includegraphics[width=1\columnwidth]{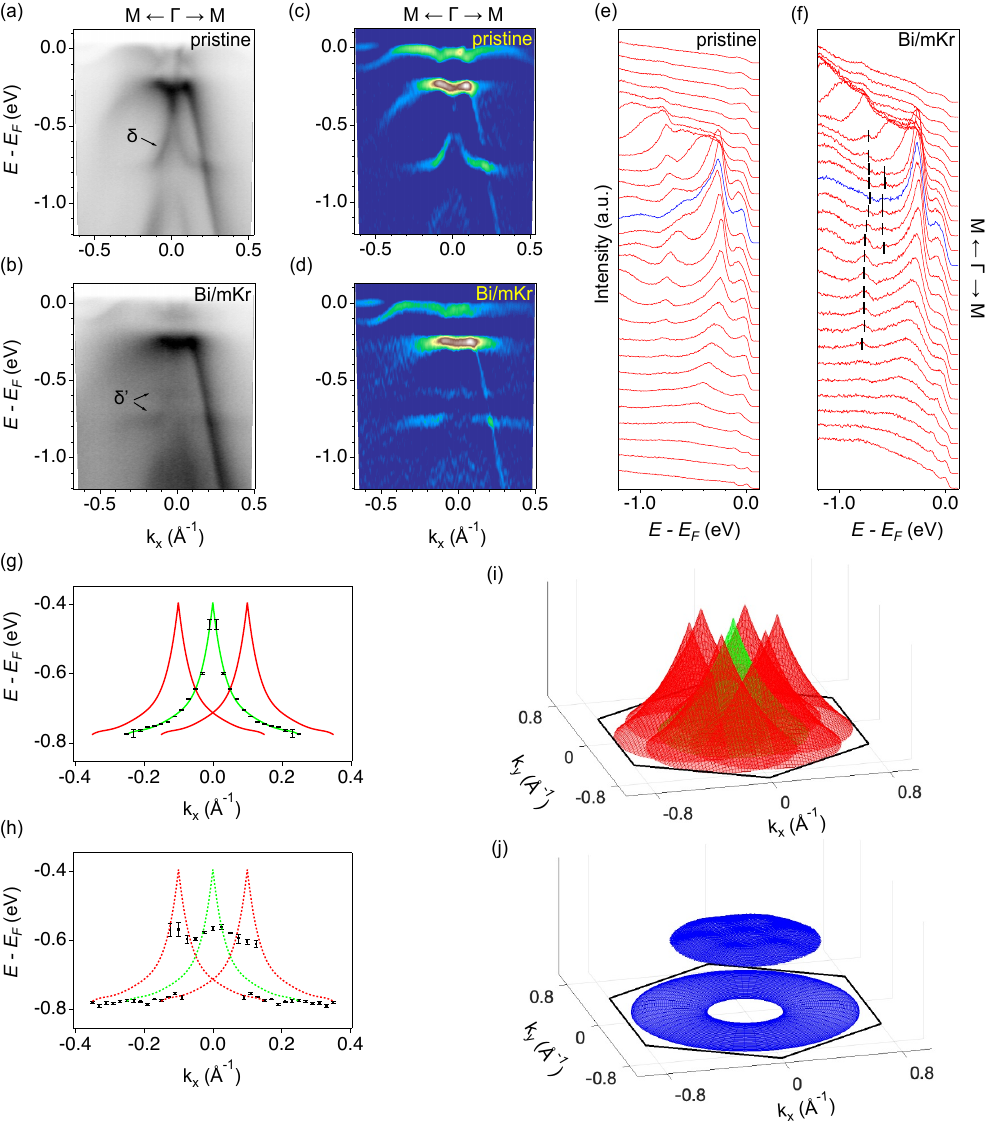}
	\caption{(a, b) Pristine Bi and Bi(111)/mKr band structure measured along $\bar{\Gamma}-\bar{M}$ direction, pristine band $\delta$ and moiré band $\delta$' are point by the black arrows.  (c, d) Curvature of a, b. (e, f) the EDC curvature of a and b, hybridization band $\delta$’ are traced by black dots.(g) Evolution of band $\delta$, black dots are EDC peak fitting,  green line and blue line are dots fitting of band $\delta$ and $\delta’$, red line is theoretical moiré band $\delta_{m}$ curve alone $\bar{\Gamma}-\bar{M}$. (h)Comprehension between theoretical moiré replicas and the real hybridization band, the red and green dash lines are same as (g) and black dots are EDC peak fitting of $\delta$' in (b). (i)Schematic of pristine and moiré replicas of band $\delta$, green part is pristine band $\delta$ and red part is theoretical moiré band  $\delta_{m}$. (j)Schematic of the Bi(111)/Kr hybridization $\delta$' band.}
\end{figure}

The moiré period in Bi/mAr is relatively small; therefore, the hybridization between the original surface states and their replicas is moderate. To achieve strong hybridization, we investigate the band structure of Bi/mKr, which has a larger moiré period ($a_{m} $= 42Å). In Figs. 4a-b, we plot band structures of pristine Bi(111) and Bi/mKr, respectively, along the $\Gamma$-M direction. We observe that the surface band $\delta$  splits into two flat branches centered at binding energies $E_{B} \sim$ eV and $E_{B} \sim$ eV following the deposition of a Kr monolayer (Fig. 4b). This band splitting feature is more clearly visualized in the second derivative plots (Figs. 4c–d) and EDC plots (Figs. 4e–f). 

The splitting of surface states is a direct consequence of band hybridization induced by the moiré effect. Compared to Bi/mAr, Bi/mKr exhibits a larger moiré period and a smaller BZ.Correspondingly, the original surface band $\delta$ and its replicas $\delta_m$ are positioned closer to each other (Figs. 4g and 4i). In contrast to Bi/mAr where the surface band and its replicas merely overlap with moderate hybridization (Fig. 3j), Bi/mKr exhibits strong band renormalization and splitting into two components, indicating band hybridization at the band crossings (Fig. 4h). This strong hybridization is unambiguously evidenced by the double-peak feature in the EDC plot of Bi(111)/mKr (Fig. 4f). Although the original surface states of Bi(111) near the $E_{F}$  exhibit complex dispersions, we also observe evidence of a strong moiré modulation effect in Bi(111)/mKr, characterized by flat-band features at binding energies $E_{B} \sim$ 0.1eV and $E_{B} \sim$ 0.25eV (Figs. 4b and 4d).

In summary, we successfully constructe mixed-dimensional Bi(111)/noble gas monolayer heterostructures. We observe that the interfacial moiré pattern modulates only the surface states, whereas the bulk states remain nearly unchanged. Additionally, the moiré period can be tuned by depositing different noble gas monolayers on the Bi(111) surface and by varying the annealing temperature. In Bi/mKr, the moiré period is relatively large, allowing us to observe hybridization between surface state replicas and signatures of moiré flat bands. Our results provide an ideal platform for investigating the transition of the moiré effect from 2D to mixed-dimensional systems, and offer insights into the electronic structure interactions within moiré patterns. Our findings are anticipated to inform further studies on moiré modulation of non-trivial surface states, as well as moiré-controlled topological phase transitions in topological insulators, topological semimetals, and topological superconductors.

We acknowledge Shengjun Yuan for useful discussions. This work was supported by the Fundamental Research Funds for the Central Universities(Grants No. 413000128), the National Natural Science Foundation of China (NSFC) (Grants No.12274329, and No. 12404083) and the China Postdoctoral Science Foundation (Grant No. 2023M732717).

\bibliography{Bismuth-2025.9.15}

\end{document}